\def\year{2016}\relax
\documentclass[letterpaper]{article}
\usepackage{aaai16}
\usepackage{times}
\usepackage{helvet}
\usepackage{courier}
\usepackage[T1]{fontenc}
\usepackage[utf8]{inputenc}
\usepackage{float}
\usepackage{calc}
\usepackage{amsmath}
\usepackage{amsthm}
\usepackage{amssymb}
\usepackage{graphicx}
\usepackage{esint,subfig}

\usepackage{xcolor}

\makeatletter

\floatstyle{ruled}
\newfloat{algorithm}{tbp}{loa}
\providecommand{\algorithmname}{Algorithm}
\floatname{algorithm}{\protect\algorithmname}

\theoremstyle{plain}
\newtheorem{thm}{\protect\theoremname}
\theoremstyle{definition}

\newtheorem*{problem}{\protect\problemname}
\ifx\proof\undefined
\newenvironment{proof}[1][\protect\proofname]{\par
\normalfont\topsep6\p@\@plus6\p@\relax
\trivlist
\itemindent\parindent
\item[\hskip\labelsep\scshape #1]\ignorespaces
}{%
\endtrivlist\@endpefalse
}
\providecommand{\proofname}{Proof}
\fi

\usepackage{xspace}

\makeatother

\providecommand{\problemname}{Problem}
\providecommand{\theoremname}{Theorem}
\begin{document}
\global\long\def\p{\operatorname{Pr}}
\global\long\def\w{\mathbf{w}}
\global\long\def\s{\mathbf{s}}
\global\long\def\d{\mathrm{d}}
\global\long\def\t{\tau}
\global\long\def\a{\mathbf{A}}
\global\long\def\c{\mathbf{C}}
\global\long\def\dd{\mathbf{d}}
\global\long\def\h{\mathbf{H}}
\global\long\def\b{\mathbf{B}}
\global\long\def\lt{\operatorname{LowerTri}}
\global\long\def\u{\mathbf{u}}
\global\long\def\s{\mathbf{S}}
\global\long\def\D{\mathbf{D}}
\global\long\def\argmax{\operatorname{arg\,max}}
\global\long\def\tt{\mathbf{t}}
\global\long\def\j{\jmath}
\global\long\def\ut{\operatorname{StrictlyUpperTri}}
\global\long\def\J{\mathbf{J}}
\global\long\def\alg{\textsc{Render}\xspace}

\title{Seeing the Unseen Network: \\
       Inferring Hidden Social Ties from Respondent-Driven Sampling}

\author{Lin Chen$^{1,2}$ \and Forrest W. Crawford$^{2,3}$ \and Amin Karbasi$^{1,2}$\\
$^1$Department of Electrical Engineering, 
$^2$Yale Institute for Network Science, 
$^3$Department of Biostatistics\\
Yale University, New Haven, CT 06520\\
\{lin.chen, forrest.crawford, amin.karbasi\}@yale.edu
}
\nocopyright
\maketitle
\begin{abstract}
Learning about the social structure of hidden and hard-to-reach 
populations --- such as drug users and sex workers --- is a major 
goal of epidemiological and public health research on risk 
behaviors and disease prevention.
Respondent-driven sampling (RDS) is a  peer-referral process widely used by many health organizations,
where research subjects recruit other subjects from their social network.
In such surveys, researchers observe who recruited whom, along with the 
time of recruitment and
the total number of acquaintances (network degree) of respondents. 
However,
due to privacy concerns, the identities of acquaintances are not disclosed.
In this work, we show how to reconstruct the underlying
network structure through which the subjects are recruited. We formulate
the dynamics of RDS as a continuous-time diffusion process over the
underlying graph and derive the likelihood of the recruitment time
series under an arbitrary inter-recruitment time distribution. 
We develop an efficient stochastic optimization algorithm called \alg (REspoNdent-Driven nEtwork Reconstruction) 
that finds the 
network that best explains the collected data. We 
support our analytical results through an exhaustive
set of experiments on both synthetic and real data. 
\end{abstract}

\section{Introduction}

Random sampling is an effective way for researchers to learn about a population of interest. Characteristics of the sample can be generalized to the population of interest using standard statistical theory. However, traditional random sampling may be impossible when the target population is hidden, e.g. drug users, men who have sex with men, sex workers and homeless people.  Due to concerns like privacy, stigmatization, discrimination, and criminalization, members of hidden populations may be reluctant to participate in a survey. In addition to the hidden populations for which no sampling ``frame'' exists, there are rare populations of great research interest, marked by their tiny fractions in the entire population. It is unlikely that random sampling from the general population would accrue a reasonable sample from a very rare population. However, it is often the case that members of hidden or rare populations know each other socially. This suggests that social referral would be an effective method for accruing a large sample. To this end, researchers have developed a variety of social link-tracing survey designs.  The most popular is \emph{respondent-driven sampling} (RDS) \cite{heckathorn1997respondent}. 

In RDS, a small set of initial subjects, known as ``seeds'' are selected, not necessarily simultaneously, from the target population. Subjects are given a few coupons, each tagged with a unique identifier, which they can use to recruit other eligible subjects.  Participants are given a reward for being interviewed and for each eligible subject they recruit using their coupons.   
Each subject reports their degree, the number of others whom they know in the target population.  
No subject is permitted to enter the study twice and the date and time of each interview is recorded. 

While RDS can be an effective recruitment method, it reveals only incomplete social network data to researchers. Any ties between recruited subjects along which no recruitment took place remain unobserved. 
The social network of recruited subjects is of great interest to sociologists, epidemiologists and public health researchers since it may induce dependencies in the outcomes of sampled individuals. 
Fortunately, RDS reveals information about the underlying social network that can be used to (approximately) reconstruct it.
By leveraging the time series of recruitments, the degrees of recruited subjects, coupon information, and who recruited whom, it is possible to interpret the induced subgraph of RDS respondents as a simple random graph model \cite{crawford2014graphical}. 

In this paper, we introduce a flexible and expressive stochastic model of RDS recruitment on a partially observed network structure.  We derive the likelihood of the observed time series; the model admits any edgewise inter-recruitment time distribution.  We propose a stochastic optimization algorithm \alg (REspoNdent-Driven nEtwork Reconstruction) to estimate unknown parameters and the underlying social network.
We conduct extensive experiments, on synthetic and real data, to confirm the accuracy and the reconstruction performance of \alg. In particular,  we apply \alg to reconstruct the network of injection drug users from an RDS study in St. Petersburg, Russia.

\section{Related Work}\label{sec:related}

RDS has been modeled as a random walk with replacement on a graph
at its equilibrium distribution \cite{heckathorn1997respondent,salganik2004sampling,volz2008probability,goel2009respondent,gile2010respondent},
and under this argument the sampling probability is proportional to
degree, which is the basis for an estimator of the population mean \cite{heckathorn2002respondent,salganik2006variance}.
In this paper, we adopt an approach that focuses on the network structure estimable from the 
RDS sample using recruitment information \cite{crawford2014graphical}. 
\citeauthor{crawford2014graphical}~(2016) assumes that edge-wise inter-recruitment times follow the exponential distribution, but this approach is 
relatively inflexible.
In many other contexts, dynamic or random processes can reveal structural information of an underlying, partially observed, network \cite{kramer2009network,shandilya2011inferring,linderman2014discovering}.
Network reconstruction and the edge prediction problem have been studied
for diffusion processes where 
multiple realizations of the process on the same
network are available \cite{liben2007link,gomez2010inferring,gomez2011uncovering}.
In the case of RDS, reconstruction is particularly challenging because only a single 
realization of the diffusion process is observed. Furthermore, we must account 
for the role of coupons in recruitment as they intricately introduce bias.  However, in contrast to  general diffusion processes over graphs, some important network topological information is 
revealed by RDS. In this study, we leverage all the available data routinely collected during
real-world RDS studies. 


\section{Problem Formulation}\label{sec:problem_formulation}
We conform to the following notation throughout the
paper. 
Suppose that $f$ is a real-valued function and that $\mathbf{v}$
is a vector. Then $f(\mathbf{v})$ is a vector of the same size as
vector $\mathbf{v}$ and the $i$-th entry of $f(\mathbf{v})$ is
denoted by $f(\mathbf{v})_{i}$, whose value is given by $f(\mathbf{v}_{i})$.
The transposes
of matrix $\a$ and vector $\mathbf{v}$ are written as $\a'$ and
$\mathbf{v}'$, respectively. And let $\mathbf{1}$ be the all-ones column vector.
\subsection{Dynamics of Respondent-Driven Sampling}

We characterize the social network of the hidden population 
as a finite undirected simple graph $G=(V,E)$ 
with no parallel edges or self-loops. Members of the hidden population are
vertices; $\{i,j\}\in E$ implies that $i\in V$ and $j\in V$ know each other.
Using RDS, researchers recruit members from the hidden
population into the study. The time-varying recruitment-induced subgraph
$\{G_{S}(t)=(V_{S}(t),E_{S}(t)):0\leq t\leq t_{F}\}$ is a nested collection
of subgraphs of $G$, where $t_{F}$ is the termination time
of the study. 
For all $0\leq t\leq t_{F}$, $G_{S}(t)$ is a subgraph of $G$. 
Here, $G_{S}(0)$ is the null graph since there are no subjects  in the study initially. 
For simplicity,
we write $G_{S}=(V_{S},E_{S})$ for $G_{S}(t_{F})=(V_{S}(t_{F}),E_{S}(t_{F}))$, and
call this the \emph{recruitment-induced subgraph} or
\emph{induced subgraph} unless we explicitly specify the time $t$.
The vertex set of the time-varying recruitment-induced subgraph at
time $t$ (i.e., $G_{S}(t)$) denotes the members in the hidden population
that are known to the study by time $t$. The subgraph $G_S(t)$ is induced by the vertex set $V_S(t)$; i.e., $E_S(t)=\{\{i,j\}|i,j\in V_S(t), \{i,j\}\in E\}$.  The time-varying recruitment-induced
subgraph evolves in the following way \cite{crawford2014graphical}.
\begin{enumerate}
\item At time $\tilde{t}$, researchers recruit a subject in the
population as a seed. This subject is included in the vertex
set $V_{S}(t)$ of $G_{S}(t)$ for all $t\geq\tilde{t}$. Researchers
may provide this subject with coupons to recruit its neighbors.
\item Once recruited into this study (either by researchers or its neighbors)
at time $\tilde{t}$, subjects currently holding coupons will attempt 
to recruit their yet-unrecruited neighbors. The inter-recruitment
time along each edge connecting a recruiter with an unrecruited neighbor 
is i.i.d. with cdf $F(t)$. Recruitment happens
when a neighbor is recruited into the
study 
and is provided with a number of coupons. A successful
recruitment costs the recruiter one coupon.
\end{enumerate}
The directed recruitment graph is $G_{R}=(V_{R},E_{R})$, where $V_{R}=V_{S}(t_{F})$
is the set of members in the study at the final stage. For two subjects
$i,j\in V_{R}$, $(i,j)\in E_{R}$ if and only if $i$ recruits $j$.
Note that the subjects recruited by researchers (i.e., the seeds)
have zero in-degree in $G_{R}$. Let $n$ denote the cardinality
of $V_{R}$ (equivalently $V_{S}(t_{F})$).
For simplicity, we label the subject recruited in the $i$th recruitment
event by $i$. The labels of the subjects in the study are $1,2,3,\ldots,n$. 
The vector of recruitment times is 
$\mathbf{t}=(t_{1},t_{2},t_{3},\ldots,t_{n})$, where $t_i$ is the recruitment time of subject $i$.
In shorthand, we write $\t(i;j)=t_{j-1}-t_{i}$ for $i<j$.
Let $\c$ be the $n\times n$ coupon matrix whose element $\c_{ij}=1$
if subject $i$ has at least one coupon just before the $j$th recruitment
event, and zero otherwise. The rows and columns of the coupon matrix
are ordered by subjects' recruitment time. The degree vector is 
$\mathbf{d}=(d_{1},d_{2},d_{3},\ldots,d_{n})'$, where $d_{i}$ is the degree of $i$ in $G$.
At time $t$ (where $t\neq t_{i}$ for $i=1,2,3,\ldots,n$), if a subject
has at least one coupon and at least one neighbor not in the study,
we term it a \emph{recruiter} at time $t$; if a subject has not entered the study
and has at least one neighbor with at least one coupon, we term it
a \emph{potential recruitee} at time $t$. 
We assume that the observed data from a RDS recruitment process consist
of $\mathbf{Y}=(G_{R},\dd,\mathbf{t},\c)$.

\subsection{Likelihood of Recruitment Time Series}

We assume that the inter-recruitment time along an edge connecting a recruiter and
potential recruitee is i.i.d. with cdf $F(t)$. Let 
$F_{s}(t)=\p\left[W\leq t\mid W>s\right]$, where $W$ is the random 
inter-recruitment time and $\p\left[W\leq t\right]=F(t)$.
We write $f_{s}(t) = F'_s(t)$ for the corresponding conditional pdf. Let $S_{s}(t)=1-F_{s}(t)$
be the conditional survival function and $H_{s}(t)=f_{s}(t)/S_{s}(t)$
be the conditional hazard function. 

We now derive a closed-form expression for the likelihood
of the recruitment time series $L(\tt|G_{S},\theta)$.
In what follows, let $M$ be the set of seeds, and let $\a$ denote the adjacency matrix representation
of the recruitment-induced subgraph at the final stage $G_{S}$; thus
we use $G_{S}$ and $\a$ interchangeably throughout this paper.
\begin{thm}
\label{thm:likelihood}\emph{\textbf{(Proof in Appendix).}} Let $R(i)$ and $I(i)$ be the recruiter set
and potential recruitee set just before time $t_{i}$, respectively, and $M$
be the set of seeds. The following statements with respect to the
likelihood of the recruitment time series hold.
\begin{enumerate}
\item The likelihood of the recruitment time series is given by
\begin{small}
\begin{eqnarray*}
&&L(\mathbf{t}|G_{S},\theta)\\&=&
\prod_{i=1}^{n}\left(\sum_{u\in R(i)}|I_{u}(i)|H_{\t(u;i)}(t_{i}-t_{u})\right)^{1\{i\notin M\}}\\
&\times&\prod_{j\in R(i)}S_{\t(j;i)}^{|I_{j}|}(t_{i}-t_{j}),
\end{eqnarray*}
\end{small}
where $\t(u;i)=t_{i-1}-t_{u}$.

\item Let $\mathbf{m}$ and $\u$ be  column vectors of size $n$ defined as $\mathbf{m}_{i}=1\{i\notin M\}$ and $\u=\dd-\a\cdot\mathbf{1}$,  and let $\h$ and $\s$ be
$n\times n$ matrices, defined as  $\h_{ui}=H_{\t(u;i)}(t_{i}-t_{u})$ and $\s_{ji}=\log S_{\t(j;i)}(t_{i}-t_{j})$. Furthermore, we form matrices $\b=(\c\circ\h)$ and $\D=(\c\circ\s)$, where $\circ$ denotes the Hadamard (entrywise) product. We let 
\begin{eqnarray*}
\beta&=&\log(\b'\u+\lt(\a\b)'\cdot\mathbf{1})\\
\delta&=&\D'\u+\lt(\a\D)'\cdot\mathbf{1},
\end{eqnarray*}
where $\lt(\cdot)$
denotes the lower triangular part (diagonal elements inclusive) of
a square matrix.  Then, the log-likelihood of the recruitment time series
can be written as
\[
l(\mathbf{t}|G_{S},\theta)=\mathbf{m}'\beta+\mathbf{1}'\delta.
\]


%
\end{enumerate}
\end{thm}

\section{Network Reconstruction Problem}\label{sec:network_recon}

Given the observed time series, we seek to reconstruct the $n\times n$
binary symmetric, zero-diagonal adjacency matrix $\a$ of $G_S$ 
and the parameter $\theta\in\Theta$ ($\Theta$ is the parameter
space) that maximizes $\p(\a,\theta|\tt)$. Recall that we use $G_{S}$ and $\a$ interchangeably throughout this paper. We have
\[
\p(\a,\theta|\tt)\propto L(\tt|\a,\theta)\p(\a,\theta),
\]
where $\p(\a,\theta)$ is the prior distribution for $(\a,\theta)$. The constraint for the parameter $\theta$
is obvious: $\theta$ must reside in the parameter space $\Theta$;
i.e., $\theta\in\Theta$. Now we discuss the constraint for the adjacency matrix $\a$---we require
that the adjacency matrix $\a$ must be \emph{compatible}.


We seek the matrix $\a$ that maximizes the probability
$\p(\a,\theta|\tt)$. 
However, we know that the directed recruitment subgraph, if viewed as an undirected
graph, must be a subgraph of the true recruitment-induced subgraph.
Let $\a_{R}$ be the adjacency matrix of the recruitment subgraph
when it is viewed as an undirected graph; i.e., the $(i,j)$ entry
of $\a_{R}$ is $1$ if a direct recruitment event occurs between
$i$ and $j$ (either $i$ recruits $j$ or $j$ recruits $i$), and
is $0$ otherwise. We require that $\a$ be greater than or equal to $\a_{R}$ entrywise.  
Recall that every subject in the study reports its degree; thus the
adjacency matrix should also comply with the degree constraints. 
Following \cite{crawford2014graphical}, we say that a symmetric, 
binary and zero-diagonal matrix $\a$ is
a \emph{compatible adjacency matrix} if 
$\a\geq\a_{R}$ entrywise, and 
$\a\cdot\mathbf{1}\leq\mathbf{d}$ entrywise.

\subsection{Problem Statement}

Now we formulate this problem as an optimization problem.
\begin{problem}
Given the observed data $\mathbf{Y}=(G_{S},\dd,\mathbf{t},\c)$, we seek
an $n\times n$ adjacency matrix $\a$ (symmetric, binary
and zero-diagonal) and a parameter value $\theta\in\Theta$ that 
\[
\begin{array}{cc}
\text{maximizes} & L(\mathbf{t}|\a,\theta)\p(\a,\theta)\\
\text{subject to} & \a\geq\a_{R}\text{ (entrywise)}\\
 & \a\cdot\mathbf{1}\leq\mathbf{d}\text{ (entrywise)} 
\end{array}
\]
\end{problem}

\subsection{Alternating Inference of $\a$ and $\theta$ \label{sub:Parameter-Estimation}}

Given the observed data $\mathbf{Y}$, we wish to infer the adjacency matrix $\a$ of the recruitment-induced graph and the distribution parameter $\theta$. 
Given $\a$, the maximum likelihood estimator (MLE) for $\theta$ is 
\[
\hat{\theta}=\argmax_{\theta\in\Theta} L(\tt|\a,\theta)\p(\a,\theta).
\]
Similarly, given the true parameter $\theta$, the MLE for the adjacency matrix $\a$ is given by
\[\hat{\a} = \argmax_{\a\mbox{ is compatible}} L(\tt|\a,\theta)\p(\a,\theta).\]
In practice, both the parameter $\theta$ and the true recruitment-induced subgraph $G_{S}$ are unknown
and need estimation. However, we can alternately estimate $\a$
and $\theta$. This is what  \alg  (presented in Algorithm~\ref{alg:Alternate-estimation-of}) does.
Each iteration is divided into two steps: $\a$-step and $\theta$-step.

\begin{algorithm}
\textbf{Input:} the observed data $\mathbf{Y}=(G_{R},\dd,\mathbf{t},\c)$;
the initial guess for the distribution parameter $\theta$, denoted
by $\hat{\theta}_{0}$; the maximum number of iterations, $\iota_{\mathrm{max}}$.

\textbf{Output:} the estimated adjacency matrix $\a$ 
(denoted by $\hat{\a}$) and
the estimated parameter $\theta$ (denoted by
$\hat{\theta}$) %

\fbox{\begin{minipage}[t]{0.97\columnwidth}%
$\iota\gets0$

\textbf{while }$\iota<\iota_{\mathrm{max}}$

\qquad{}$\hat{\a}_{\iota}\gets\argmax_{\a\text{ is compatible}}L(\mathbf{t}|\a,\hat{\theta}_{\iota})\p(\a,\hat{\theta}_{\iota})$\hfill{}($\a$-step, we use Algorithm~\ref{alg:SA} here.)

\qquad{}$\hat{\theta}_{\iota+1}\gets\argmax_{\theta\in\Theta}L(\tt |\hat{\a}_{\iota},\theta)\p(\hat{\a}_{\iota},\theta)$\hfill{}($\theta$-step)

\qquad{}$\iota\gets\iota+1$

\textbf{end while}

$\hat{\a}\gets\hat{\a}_{\iota_{\mathrm{max}}-1}$

$\hat{\theta}\gets\hat{\theta}_{\iota_{\mathrm{max}}}$%
\end{minipage}}\caption{\alg: Alternating inference of $G_{S}$ and $\theta$ \label{alg:Alternate-estimation-of}}
\end{algorithm}

\subsubsection{Estimation of $\a$ using simulated annealing}

The $\a$-step of \alg 
solves 
\[
\max_{\a\text{ is compatible}}L(\tt|\a,\hat{\theta}_{\iota})\p(\a,\hat{\theta}_{\iota});
\]
equivalently, it suffices to solve 
\[
\max_{\a\text{ is compatible}}(l(\tt | \a,\hat{\theta}_{\iota})+\log\p(\a, \hat{\theta}_{\iota})).
\]
We employ a simulated-annealing-based method to estimate the adjacency
matrix $\a$ (presented in Algorithm~\ref{alg:SA}). Let the energy function be
\[
\Lambda_{\gamma}(\a;\tt,\hat{\theta}_{\iota})\triangleq\exp\left[-\left(l(\tt | \a,\hat{\theta}_{\iota})+\log\p(\a,\hat{\theta}_{\iota})\right)/\gamma\right],
\]
where $\gamma$ is the \emph{temperature}. 
We specify a cooling schedule, which
is a sequence of positive numbers $\{\gamma_{\j}\}_{\j\geq1}$ that
satisfy $\lim_{\j\to\infty}\gamma_{\j}=0$, where $\gamma_{\j}$ is
the temperature in the $\j$-th iteration. Note that the $\jmath$-th iteration of the simulated annealing procedure
has a compatible adjacency matrix $\a(\jmath)$ as its
state.
\begin{algorithm}[tbh]
\textbf{Input: }the number of\textbf{ }iterations, $\j_{\mathrm{max}}$;
the cooling schedule $\{\gamma_{\jmath}\}_{\j\geq1}$; initial compatible
adjacency matrix $\a(1)$; estimated parameter $\hat{\theta}_{\iota}$. 

\textbf{Output:} the estimated adjacency matrix $\hat{\a}_{\iota}$

\fbox{\begin{minipage}[t]{0.97\columnwidth}%
\textbf{for }$\j=1$ \textbf{to }$\j_{\mathrm{max}}$ \textbf{do}

\qquad{}Use Algorithm \ref{alg:proposal} to propose a compatible
adjacency matrix $\tilde{\a}(\j+1)$ based on $\a(\j)$.

\qquad{}$\psi\gets\min\left\{ 1,\frac{\Lambda_{\gamma_{\j}}(\tilde{\a}(\j+1);\tt,\hat{\theta}_{\iota})}{\Lambda_{\gamma_{\j}}(\a(\j);\tt,\hat{\theta}_{\iota})}\cdot\frac{\p(\a(\jmath)|\tilde{\a}(\jmath+1))}{\p(\tilde{\a}(\jmath+1)|\a(\jmath))}\right\} .$

\qquad{}$\a(\j+1)\gets\begin{cases}
\tilde{\a}(\j+1) & \mbox{with probability }\psi;\\
\a(\j) & \mbox{with probability }1-\psi.
\end{cases}$

\textbf{end for}

$\hat{\a}_{\iota}\gets\a(\j_{\mbox{max}}+1)$.%
\end{minipage}}\caption{Simulated-annealing-based optimization \label{alg:SA}}
\end{algorithm}
Algorithm~\ref{alg:proposal} specifies which state (compatible
adjacency matrix) the algorithm should transition into in the next
iteration. 
Concretely, in each iteration of Algorithm~\ref{alg:SA}, it randomly proposes an edge that connects
vertices $i$ and $j$. If the edge does not appear in $\a(\jmath)$
and it still conforms to the
degree constraint if we add the edge, then we simply add it to $\tilde{\a}(\jmath+1)$. In contrast,
if the edge appears in $\a(\jmath)$ 
and it still conforms to the subgraph constraint if we remove the
edge, then we simply remove it from $\tilde{\a}(\jmath+1)$. If neither condition
is satisfied, the algorithm tries again with a new proposal. We prove in the appendix that the space of compatible adjacency matrices
is connected by the proposal method.


\begin{algorithm}[tbh]
\textbf{Input: }the compatible adjacency matrix $\a(\jmath)$ in the
$\jmath$-th iteration.

\textbf{Output:} a compatible adjacency matrix $\tilde{\a}(\jmath+1)$,
which will be a candidate for the state in the $(\jmath+1)$-th iteration.

\fbox{\begin{minipage}[t]{0.97\columnwidth}%
\textbf{loop}

\qquad{}$i,j\gets\text{two distinct random integers in }[1,n]\cap\mathbb{N}$

\qquad{}\textbf{if }$\a_{ij}(\jmath)=0$ and $\sum_{k=1}^{n}\a_{ik}(\jmath)<\dd_{i}$
and $\sum_{k=1}^{n}\a_{jk}(\jmath)<\dd_{j}$ \textbf{then}

\qquad{}\qquad{}$\tilde{\a}^{+}(\jmath+1)\gets\a(\jmath)$

\qquad{}\qquad{}$\tilde{\a}_{ij}^{+}(\jmath+1)\gets1$ and $\tilde{\a}_{ji}^{+}(\jmath+1)\gets1$


\qquad{}\qquad{}\textbf{$\tilde{\a}(\jmath+1)\gets\tilde{\a}^{+}(\jmath+1)$} and \textbf{exit loop}

\qquad{}\textbf{else}

\qquad{}\qquad{}\textbf{if} $\a_{ij}(\jmath)=1$ and $\a^{(ij)}_{R}=0$
($\a^{(ij)}_{R}$ is the $(i,j)$-entry of the matrix $\a_{R}$) \textbf{then}

\qquad{}\qquad{}\qquad{}$\tilde{\a}^{-}(\jmath+1)\gets\a(\jmath)$

\qquad{}\qquad{}\qquad{}$\tilde{\a}_{ij}^{-}(\jmath+1)\gets0$ and $\tilde{\a}_{ji}^{-}(\jmath+1)\gets0$


\qquad{}\qquad{}\qquad{}$\tilde{\a}(\jmath+1)\gets\tilde{\a}^{-}(\jmath+1)$ and \textbf{exit loop}

\qquad{}\qquad{}\textbf{end if}

\qquad{}\textbf{end if}

\textbf{end loop}

\textbf{return $\tilde{\a}(\jmath+1)$}%
\end{minipage}}\caption{Proposal of compatible adjacency matrix \label{alg:proposal}}
\end{algorithm}

The proposed compatible adjacency matrix $\tilde{\a}(\jmath+1)$ is
accepted as the state for the next iteration with probability $\psi$,
where  $\psi$ equals
\[
\min\left\{ 1,\frac{\Lambda_{\gamma_{\j}}(\tilde{\a}(\j+1);\tt,\hat{\theta}_{\iota})}{\Lambda_{\gamma_{\j}}(\a(\j);\tt,\hat{\theta}_{\iota})}\cdot\frac{\p(\a(\jmath)|\tilde{\a}(\jmath+1))}{\p(\tilde{\a}(\jmath+1)|\a(\jmath))}\right\} ;
\]
otherwise, the matrix $\a(\jmath)$ remains the state for the next
iteration. The term $\frac{\Lambda_{\gamma_{\j}}(\tilde{\a}(\j+1);\tt,\hat{\theta}_{\iota})}{\Lambda_{\gamma_{\j}}(\a(\j);\tt,\hat{\theta}_{\iota})}$
is called the \emph{likelihood ratio }and the term $\frac{\p(\a(\jmath)|\tilde{\a}(\jmath+1))}{\p(\tilde{\a}(\jmath+1)|\a(\jmath))}$
is called the \emph{proposal ratio}. To implement Algorithm~\ref{alg:SA}, we have to compute the likelihood ratio and the proposal
ratio efficiently. In fact, they can be evaluated efficiently in a recursive manner.
\begin{thm}
\label{thm:The-proposal-ratio} \emph{\textbf{(Proof in Appendix).}} The proposal ratio $\frac{\p(\a(\jmath)|\tilde{\a}(\jmath+1))}{\p(\tilde{\a}(\jmath+1)|\a(\jmath))}$
is given by
\[
\frac{\mathrm{Add}(\a(\jmath))+\mathrm{Remove}(\a(\jmath))}{\mathrm{Add}(\tilde{\a}(\jmath+1))+\mathrm{Remove}(\tilde{\a}(\jmath+1))},
\]
where 
$\mathrm{Add}(\a)=\sum_{1\leq i<j\leq n}1\{\a_{ij}=0,
\sum_{k=1}^{n}\a_{ik}<\dd_{i},\sum_{k=1}^{n}\a_{jk}<\dd_{j}\},$
%
%
and $\mathrm{Remove}(\a)=\sum_{1\leq i<j\leq n}1\{\a_{ij}=1\mbox{ and }\a_{R}^{(ij)}=0\}$. Here, we let 
$\a_{R}^{(ij)}$ denote the $(i,j)$-entry of the matrix $\a_{R}$. 
\end{thm}
The same way, we can find the likelihood ratio in a recursive manner.
\begin{thm}
\label{thm:likelihood_ratio} \emph{\textbf{(Proof in Appendix).}} 
If we view $\beta$ in Theorem \ref{thm:likelihood}
as a function of the adjacency matrix $\a$, denoted by $\beta(\a)$, then the recurrence relation between $\beta(\tilde{\a}(\j+1))$ and $\beta(\a(\j))$ is as follows: 
\begin{multline}
(e^{\beta(\tilde{\a}(\j+1))}-e^{\beta(\a(\j))})_{j}=\\
\pm\left(\b_{bj}1\{a<j\}-\b_{aj}1\{b<j\}\right).
\end{multline}
Here, we assign the minus sign ``$-$'' in ``$\pm$'' if $\tilde{\a}(\j+1)=\tilde{\a}^{+}(\jmath+1)$, and assign the plus sign ``$+$'' in ``$\pm$'' if $\tilde{\a}(\j+1)=\tilde{\a}^{-}(\jmath+1)$.
%
 By the same convention, the  likelihood ratio $\Lambda_{\gamma_{\j}}(\tilde{\a}(\j+1)|{\tt,\hat{\theta}_{\iota})} / \Lambda_{\gamma_{\j}}(\a(\j)|\tt,\hat{\theta}_{\iota})$ is given by 
\begin{multline*}
\exp\left\{ -\gamma_{\j}^{-1}\left[-\log\frac{\p(\tilde{\a}(\j+1))}{\p(\a(\j))}+\mathbf{m}'\left(\beta(\tilde{\a}(\j+1))\right.\right.\right.\\
\left.\left.\left.-\beta(\a(\j))\right)\pm\sum_{j=1}^{n}\left(\D_{bj}1\{a<j\}-\D_{aj}1\{b<j\}\right)\right]\right\}, 
\end{multline*}
\end{thm}
\subsubsection{Estimation of distribution parameter}\label{sub:para-est}

In a $\theta$-step in Algorithm \ref{alg:Alternate-estimation-of},
we have to solve the optimization problem 
$
\hat{\theta}_{\iota+1}\gets\argmax_{\theta\in\Theta}L(\theta|\hat{\a}_{\iota},\mathbf{t})\p(\hat{\a}_{\iota},\theta)
$. 
If the parameter
space $\Theta$ is a subset of the $p$-dimensional Euclidean space
$\mathbb{R}^{p}$, this problem can be solved using off-the-shelf
solvers, e.g., $\textsc{fminsearch}$ in MATLAB.

\section{Experiments}\label{sec:experiment}

\begin{figure}
  \fbox{\includegraphics[width=0.46\columnwidth]{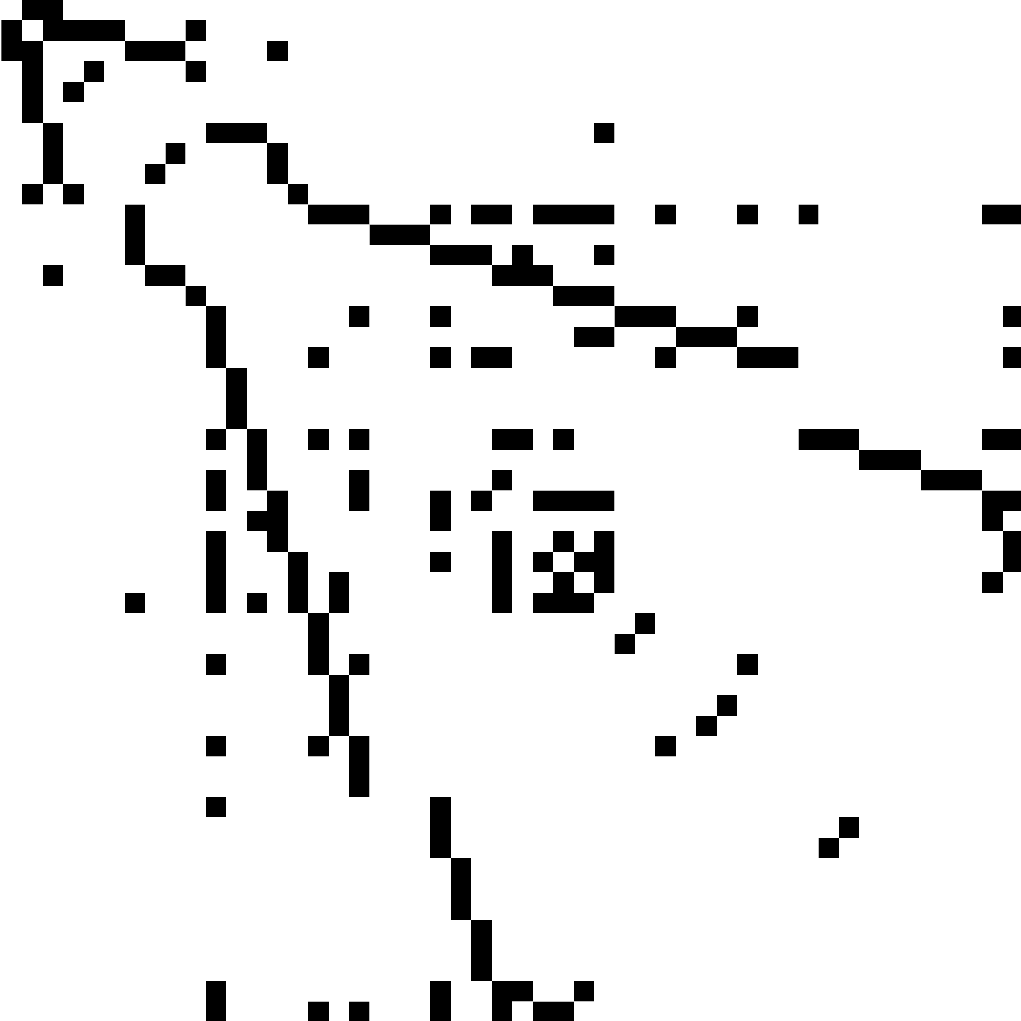}}
  \fbox{\includegraphics[width=0.46\columnwidth]{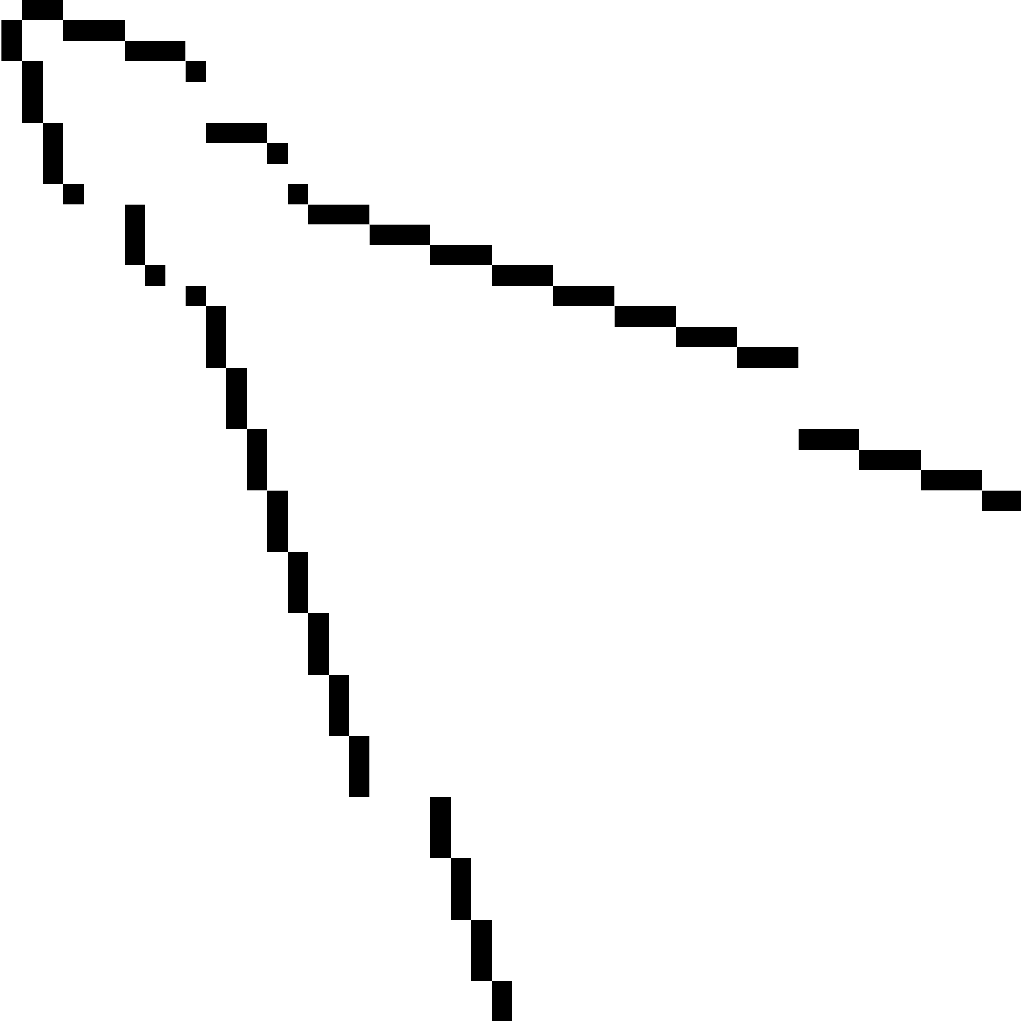}} \\
  \fbox{\includegraphics[width=0.46\columnwidth]{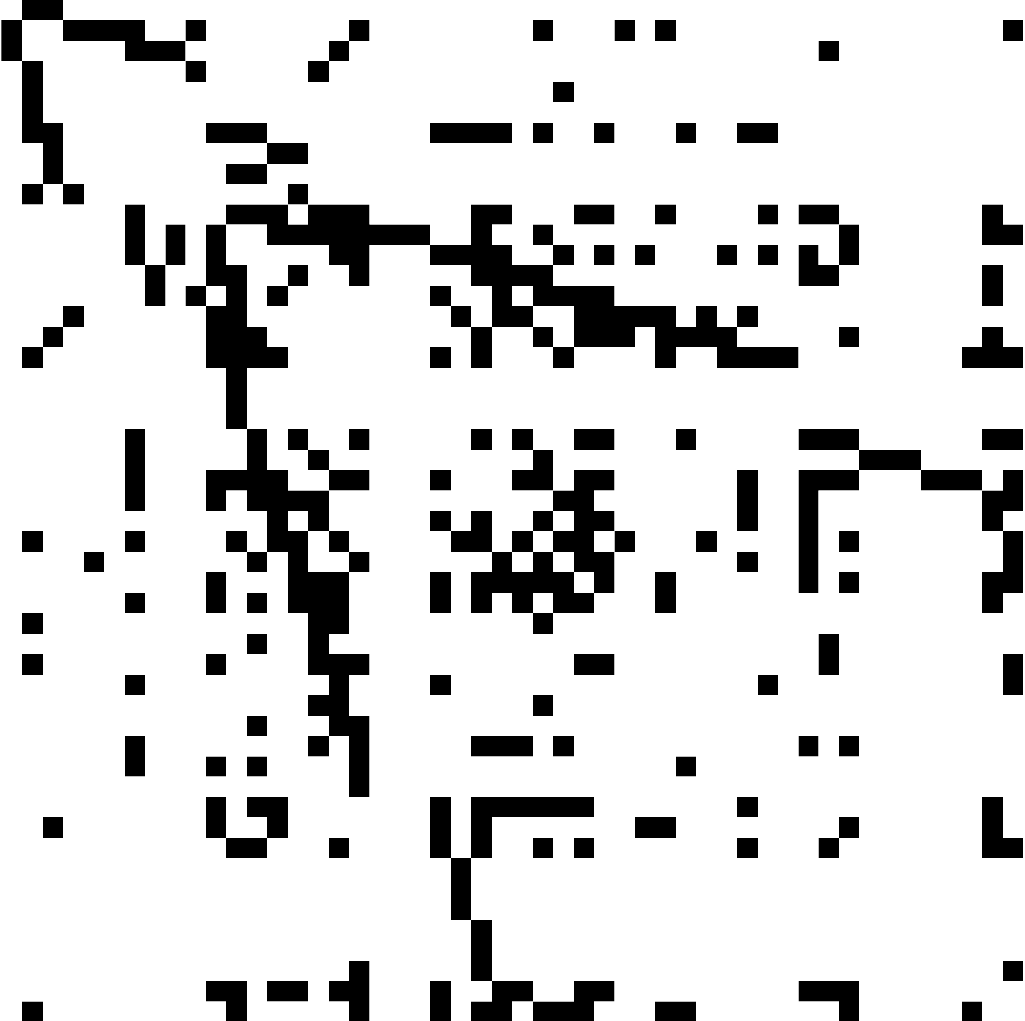}}
        \includegraphics[width=0.46\columnwidth]{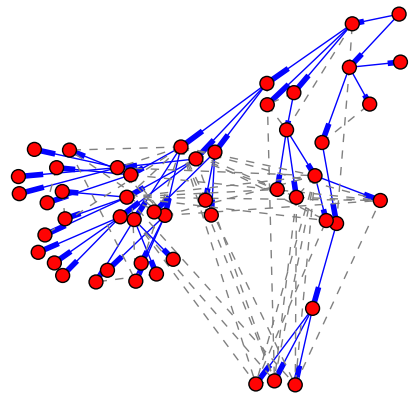}
        \caption{Example reconstruction procedure. Clockwise from top left: the true recruitment-induced subgraph $G_S$, the observed recruitment graph $G_R$, the estimated network with recruitments as blue arrows and dashed lines as inferred edges, and the estimated recruitment-induced subgraph $\hat{G}_S$.\label{fig:example}}
\end{figure}

We evaluated the proposed method in two aspects, the reconstruction
performance of the recruitment-induced subgraph and the parameter
estimation of the edgewise inter-recruitment time model. Let $\a$
be the adjacency matrix of the true recruitment-induced subgraph $G_{S}$
and $\hat{\a}$ be the estimate. We define the true and false positive rates (TPR and FPR) as $\mathrm{TPR}(\hat{\a},\a)=\sum_{i<j}1\{\hat{\a}_{ij}=1\mbox{ and }\a_{ij}=1\}/\binom{n}{2}$ and $\mathrm{FPR}(\hat{\a},\a)=\sum_{i<j}1\{\hat{\a}_{ij}=1\mbox{ and }\a_{ij}=0\}/\binom{n}{2}.$
Fig.~\ref{fig:example}  illustrates
an example of the reconstruction procedure. We simulated a RDS
process over the Project 90 graph \cite{woodhouse1994mapping}
with power-law edgewise inter-recruitment time distribution, whose
shape parameter $\alpha=2$ and scale parameter $x_{\mathrm{min}}=0.5.$
Fifty subjects are recruited in this process. 
We show clockwise from top left: the true recruitment-induced subgraph $G_S$, the observed recruitment graph $G_R$, the estimated network with recruitments as blue arrows and dashed lines as inferred edges, and the estimated recruitment-induced subgraph $\hat{G}_S$.
The TPR equals $0.769$ and the FPR equals $0.106$. The estimated
parameter $\hat{\alpha}=1.95$ and $\hat{x}_{\mathrm{min}}=0.49$.

\subsection{Reconstruction Performance}
\begin{figure}

\centering
\includegraphics[width=0.9\columnwidth]{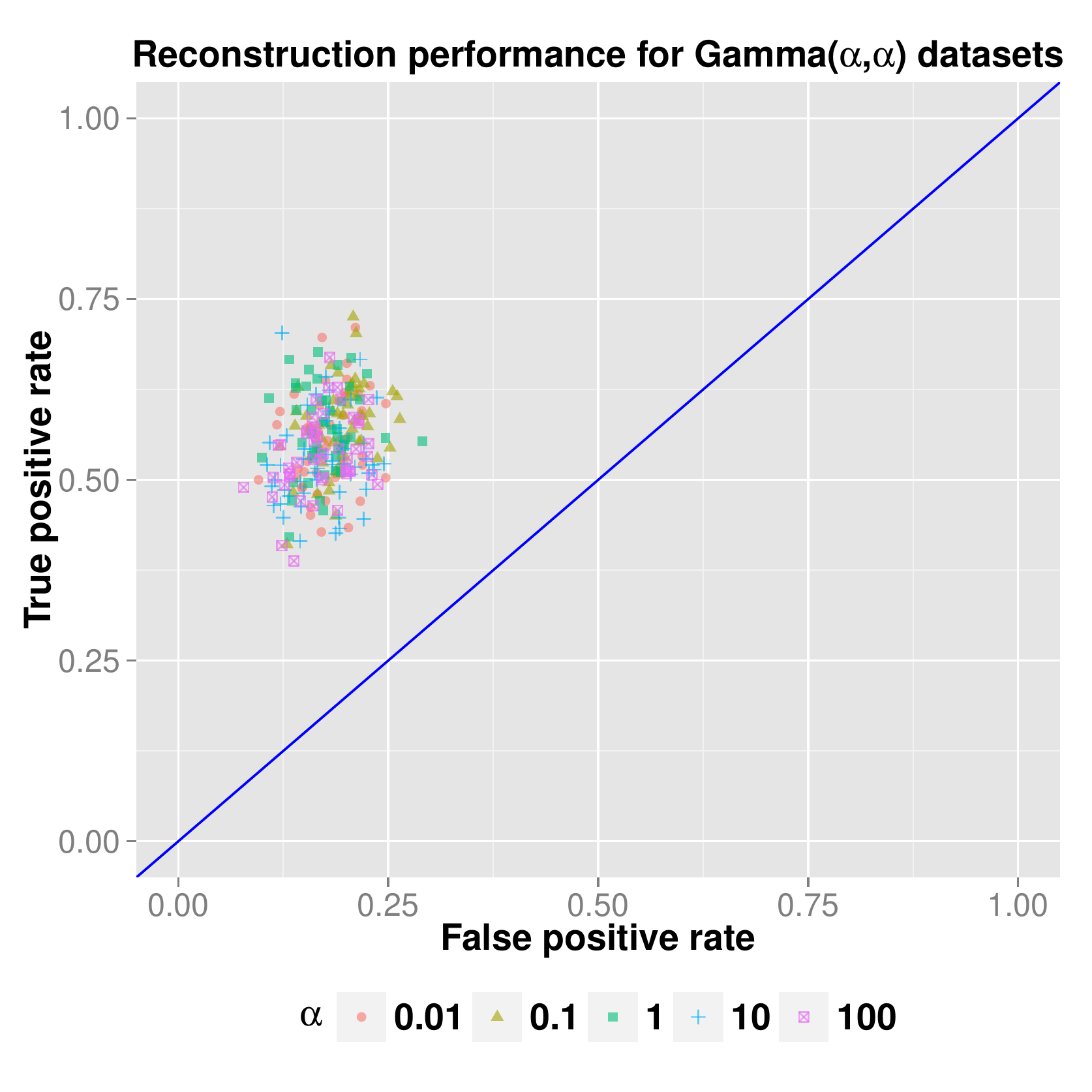}

\caption{True and false positive rates of the reconstruction results for $\mathrm{Gamma}(\alpha,\alpha)$ datasets. Each point corresponds to the reconstruction
result of one dataset. The different shapes of the points indicate different
shape parameters.\label{fig:gamma}}
\end{figure}

\begin{figure}[ht!]
\centering
 \includegraphics[width=0.9\columnwidth]{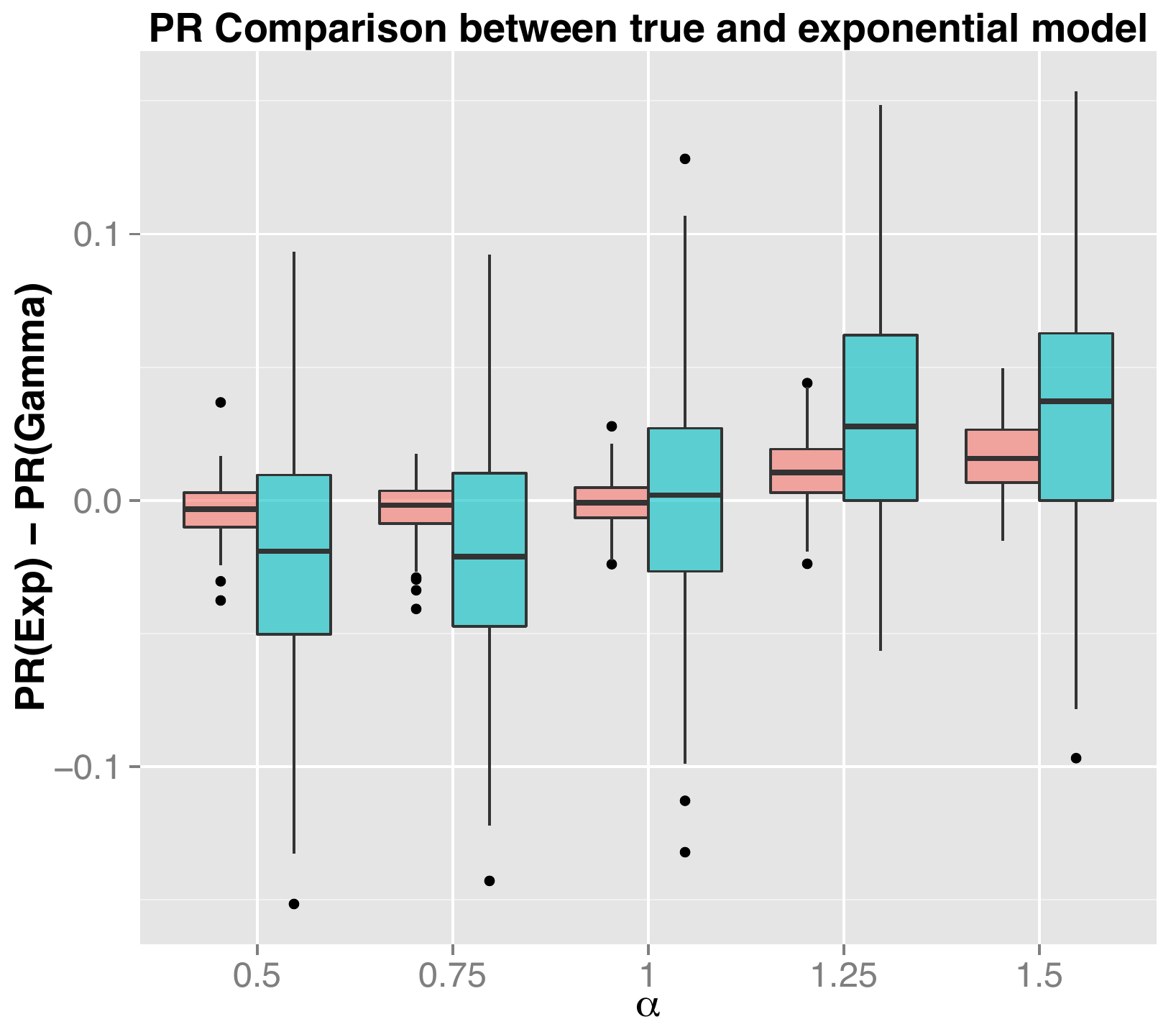}\caption{Difference between true and false positive rates (green and red boxplots, respectively) under the exponential model and the true Gamma model (the vertical axis)
with the true shape parameter $\alpha$ (the horizontal axis).
\label{fig:Comparison}}
\end{figure}

\begin{figure}[ht!]
\centering
\includegraphics[width=0.9\columnwidth]{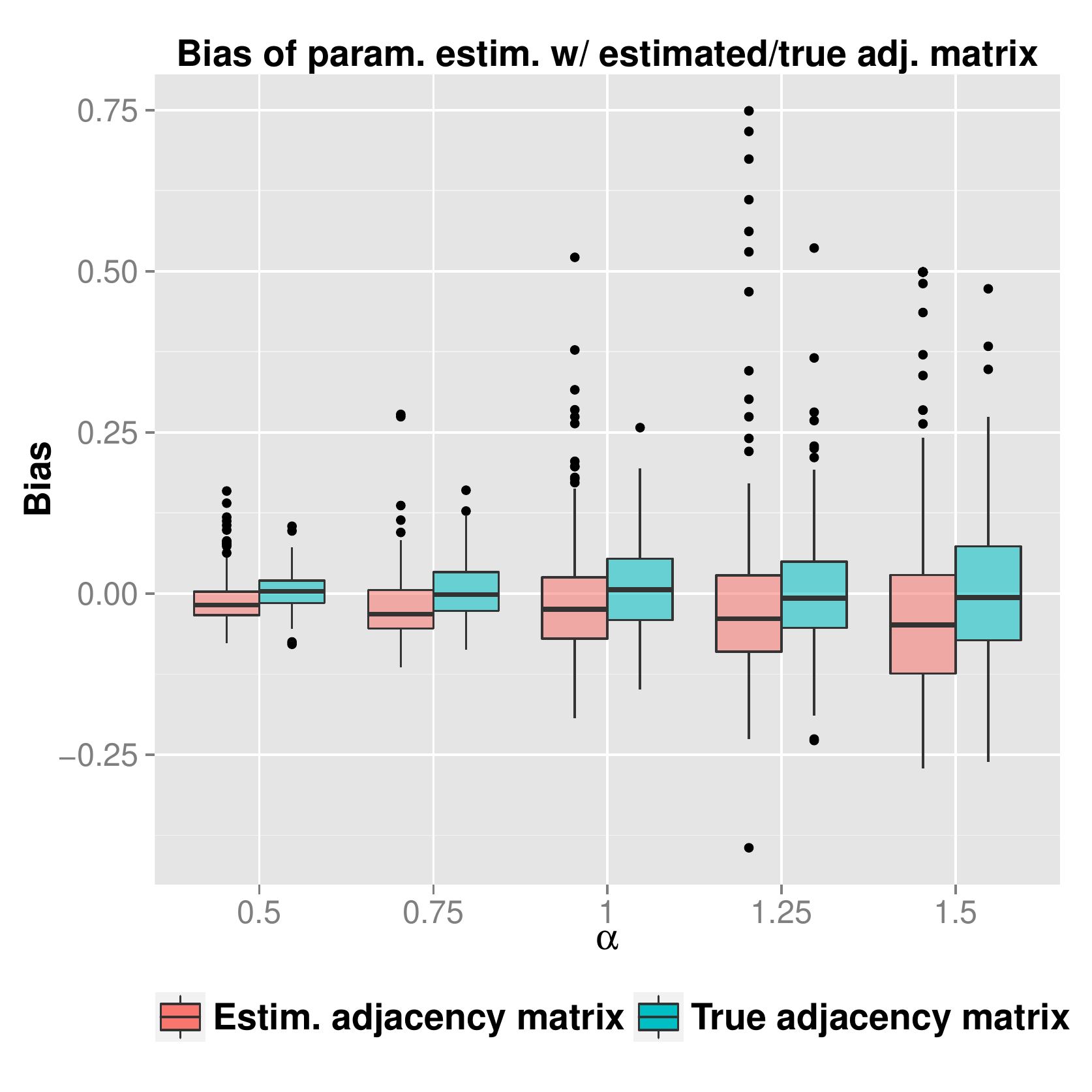}\caption{Distribution of bias in the estimated
shape parameter $\hat{\alpha}$ 
given the estimated adjacency matrix (red boxes) and the
true adjacency matrix (green boxes). The horizontal axis represents
the values of the true shape parameter $\alpha$. 
Whisker lengths are $1.5$ times the inter-quartile range. 
\label{fig:Accuracy}}
\end{figure}

\subsubsection{Impact of distribution parameter}

We simulated 50 RDS over the Project 90 graph with inter-recruitment time distribution
$\textrm{Gamma}(\alpha,\alpha)$ (parametrized by the shape and scale) 
for each $\alpha=0.01$, $0.1$, $1,$ $10,$ and $100$.
Thus a total number of 250 RDS processes are
simulated and the mean inter-recruitment time is fixed to be $1$.
The reconstruction performance is illustrated in Fig. \ref{fig:gamma}.
Each point on the receiver operating characteristic (ROC) plane represents
a reconstruction accuracy performance of a simulated RDS
process. Points with the same marker have the same inter-recruitment
time distribution parameter. From the figure, we can observe that
there is no significant sign of separation of points with different
inter-recruitment time distribution parameters. Reconstruction accuracy
is robust to the distribution parameter.

\subsubsection{Impact of distribution model}

We simulated 50 RDS
process over the Project 90 graph with inter-recruitment time distribution
$\textrm{Gamma}(0.5,0.5)$. The recruitment-induced subgraph is reconstructed
via the model of the true inter-recruitment distribution $\textrm{Gamma}(0.5,0.5)$
and the exponential distribution $\textrm{Exp}(1)$, respectively.
The TPR and FPR of each dataset are
presented in Fig. \ref{fig:Comparison}. The TPR is always higher than the FPR,
which reaffirms the effectiveness of our proposed reconstruction method.
For each dataset, the TPR and FPR under the true and the exponential models are very close to
each other. We observe that there is no significant reconstruction
skewness incurred by mis-specification of the inter-recruitment time
distribution model.

\subsection{Parameter Estimation}\label{sec:para-est}

We simulated 200 RDS  
processes over the Project 90 graph with edgewise inter-recruitment
time distribution $\textrm{Gamma}(\alpha,\alpha)$ (parametrized by
the shape and scale parameters) for each $\alpha=0.5$, $0.75$, $1,$
$1.25,$ and $1.5$. We used the method in the ``Estimation of distribution parameter'' section. 
We 
assess the bias of the estimated shape
parameter $\hat{\alpha}$, which is given by $\hat{\alpha}-\alpha$.
Fig. \ref{fig:Accuracy} shows the distribution of the
bias using Tukey boxplots. 
In Fig. \ref{fig:Accuracy}, the red boxes depict the distribution of
the biases of the estimated shape parameters inferred through the
estimated adjacency matrix, while the green boxes illustrate the distribution
of 
those
 inferred given the
true adjacency matrix.
The horizontal axis shows the value of the true shape parameter. 

With respect to the red boxes (those based on the estimated adjacency matrices), The middle line of each box is very close to zero and thus the estimation is
highly accurate. The interquartile range (IQR)
, which 
measures the deviation of the
biases, declines as the shape parameter decreases. 
Even for the box with largest deviation (i.e.,
the box with $\alpha=1.5$), the IQR is approximately $[-0.125,0.02]$
and $99.3\%$ of the biases reside in the interval $[-0.27,0.25]$.
Compared with the parameter estimation via the true adjacency matrix,
this estimator based on the estimated adjacency matrix is biased to
some degree. In Fig. \ref{fig:Accuracy}, we can observe that it underestimates
$\alpha$.


Then consider the green boxes (those based on the true adjacency matrix). 
Similar to those based on the estimated adjacency matrix, 
the deviation of this estimator declines as the
value of the shape parameter $\alpha$ decreases. The middle line of each box is noted to coincide perfectly with zero bias line, which suggests that this estimator is unbiased
given the true adjacency matrix.

\subsection{Experiments on Real Data}

\begin{figure}[b] 
\centering \includegraphics[width=0.6\columnwidth]{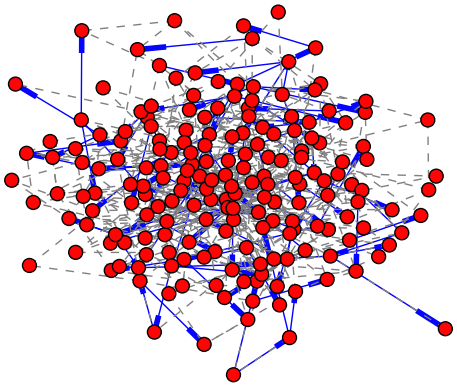}
\caption{Illustration of the largest community of the inferred underlying social
network of the St. Petersburg dataset. The blue arrows represent the
edges in the recruitment subgraph that indicates the recruiter and
recruitee. Gray dashed edges are inferred from the data. \label{fig:illu-real}}
\end{figure}

We also apply $\alg$ to data from an RDS study of
$n=813$ drug users in St. Petersburg, Russian Federation. 
We use $\alg$ to infer the
underlying social network structure of the drug users in this study.
Since it could be confusing to visualize the whole inferred network,
we only show the inferred subgraph of the largest community of the
network, as presented in Fig. \ref{fig:illu-real}. The blue arrows
represent the edges in the recruitment subgraph that indicates the
recruiter and recruitee. Gray dashed edges are inferred from the data.

\section{Conclusion}\label{sec:conclusion}
In this paper, we precisely formulated the dynamics of RDS as a continuous-time
diffusion process over the underlying graph. We derived the likelihood
for the recruitment time series under an arbitrarily recruitment time
distribution. As a result, we develop an efficient stochastic optimization 
algorithm, \alg, that identifies the optimum network that
best explains the collected data. We then supported the performance of \alg  through an exhaustive set of experiments on both synthetic
and real data. 

\section{Acknowledgements}

FWC was
supported by NIH/NCATS grant KL2 TR000140 and NIMH grant P30MH062294. LC would like to thank Zeya Zhang and Yukun Sun for their encouragement.

\bibliographystyle{aaai}
\bibliography{reference-list}


\clearpage{}
\section*{Appendix}

\subsection*{Proof of Theorem \ref{thm:likelihood}}

\begin{proof}
(1) We consider the recruitment of subject $i$. Let $R_{u}(i)$ be
the set of recruiters of subject $u$ just before time $t_{i}$ and
$I_{u}(i)$ be the set of recruitees of recruiter $u$ just before
time $t_{i}$. Suppose that $i\notin M$. The inter-recruitment time
between $i$ and its potential recruiter $u$ is denoted $W_{ui}=t_{i}-t_{u}$
and is greater than $t_{i-1}-t_{u}$ conditional on previous recruitment
of $i$. Let $U$ be the random variable of next recruiter and $X$
be the random variable of next recruitee. Let $J$ denote the event
$\forall j\in R(i),k\in I(i),W_{jk}>t_{i-1}-t_{j}$. We have
\begin{align*}
 & \p\left[U=u,X=x,W_{ux}\geq t-t_{u}\mid J\right]\\
= & \p\left[W_{ux}\geq t-t_{u},t_{j}+W_{jk}>t_{u}+W_{ux},\right.\\
 & \left.\forall j\in R(i),k\in I(i),\{u,x\}\neq\{j,k\}\mid J\right],
\end{align*}
and
\begin{eqnarray*}
 &  & \p\left[W_{Ui}\geq t-t_{U}\mid J\right]\\
 & = & \sum_{x\in I(i)}\sum_{u\in R_{x}(i)}\p\left[W_{ux}\geq t-t_{u},t_{j}+W_{jk}>t_{u}\right.\\
 &  & \left.+W_{ux},\forall j\in R(i),k\in I(i),\{u,x\}\neq\{j,k\}\mid J\right]\\
 & = & \sum_{x\in I(i)}\sum_{u\in R_{x}(i)}\int_{t-t_{u}}^{\infty}f_{\t(u;i)}(s)\d s\p\left[W_{jk}>s+t_{u}\right.\\
 &  & \left.-t_{j},\forall j\in R(i),k\in I(i),\{u,x\}\neq\{j,k\}\mid J\right]\\
 & = & \sum_{x\in I(i)}\sum_{u\in R_{x}(i)}\int_{t-t_{u}}^{\infty}f_{\t(u;i)}(s)\cdot\\
 &  & \frac{\prod_{j\in R(i)}\prod_{k\in I_{j}(i)}\left(1-F_{\t(j;i)}(s+t_{u}-t_{j})\right)}{1-F_{\t(u;i)}(s)}\d s\\
 & = & \sum_{x\in I(i)}\sum_{u\in R_{x}(i)}\int_{t-t_{u}}^{\infty}H_{\t(u;i)}(s)\cdot\\
 &  & \prod_{j\in R(i)}S_{\t(j;i)}^{|I_{j}(i)|}(s+t_{u}-t_{j})\d s
\end{eqnarray*}
Thus the likelihood is
\begin{align*}
 & \sum_{x\in I(i)}\sum_{u\in R_{x}(i)}H_{\t(u;i)}(t_{i}-t_{u})\prod_{j\in R(i)}S_{\t(j;i)}^{|I_{j}(i)|}(t_{i}-t_{j})\\
= & \prod_{j\in R(i)}S_{\t(j;i)}^{|I_{j}(i)|}(t_{i}-t_{j})\sum_{u\in R(i)}|I_{u}(i)|H_{\t(u;i)}(t_{i}-t_{u})
\end{align*}
Now suppose that $i\in M$. 
\begin{align*}
 & \p\left[W_{jk}\geq t-t_{j},\forall j\in R(i),k\in I_{j}(i)\mid J\right]\\
= & \prod_{j\in R(i)}\prod_{k\in I_{j}(i)}(1-F_{\t(j;i)}(t-t_{j}))\\
= & \prod_{j\in R(i)}S_{\t(j;i)}^{|I_{j}(i)|}(t-t_{j}).
\end{align*}
Thus the likelihood is 
\[
\prod_{j\in R(i)}S_{\t(j;i)}^{|I_{j}(i)|}(t_{i}-t_{j}).
\]
Therefore the entire likelihood is
\begin{multline*}
\prod_{i=1}^{n}\left(\sum_{u\in R(i)}|I_{u}(i)|H_{\t(u;i)}(t_{i}-t_{u})\right)^{1\{i\notin M\}}\cdot\\
\prod_{j\in R(i)}S_{\t(j;i)}^{|I_{j}|}(t_{i}-t_{j}).
\end{multline*}

(2) The log-likelihood is
\begin{multline*}
\sum_{i=1}^{n}\left[1\{i\notin M\}\log\left(\sum_{u\in R(i)}|I_{u}(i)|H_{\t(u;i)}(t_{i}-t_{u})\right)\right.\\
\left.+\sum_{j\in R(i)}|I_{j}(i)|\log S_{\t(j;i)}(t_{i}-t_{j})\right].
\end{multline*}
And the number of recruitees of recruiter $u$ just before time $t_{i}$
is given by

\[
|I_{j}(i)|=\c_{ji}\left(\sum_{k=i}^{n}\a_{jk}+\u_{j}\right).
\]
The term $\sum_{u\in R(i)}|I_{u}(i)|H_{\t(u;i)}(t_{i}-t_{u})$ in
the log-likelihood is given by

\begin{align*}
 & \sum_{u\in R(i)}|I_{u}(i)|H_{\t(u;i)}(t_{i}-t_{u})\\
= & \sum_{u}\c_{ui}\left(\sum_{k=i}^{n}\a_{uk}+\u_{u}\right)\h_{ui}\\
= & (\b'\u+\lt(\a\b)'\cdot\mathbf{1})_{i}.
\end{align*}
The term $\sum_{j\in R(i)}|I_{j}(i)|\log S_{\t(j;i)}(t_{i}-t_{j})$
in the log-likelihood is given by

\begin{align*}
 & \sum_{j\in R(i)}|I_{j}(i)|\log S_{\t(j;i)}(t_{i}-t_{j})\\
= & \sum_{j\in R(i)}\c_{ji}\left(\sum_{k=i}^{n}\a_{jk}+\u_{j}\right)\s_{ji}\\
= & (\D'\u+\lt(\a\D)'\cdot\mathbf{1})_{i}.
\end{align*}

Thus the log-likelihood is
\begin{multline*}
\sum_{i=1}^{n}\left[1\{i\notin M\}\log\left(\b'\u+\lt(\a\b)'\cdot\mathbf{1}\right)_{i}\right.\\
\left.+\left(\D'\u+\lt(\a\D)'\cdot\mathbf{1}\right)_{i}\right]=\mathbf{m}'\beta+\mathbf{1}'\delta.
\end{multline*}

\end{proof}

\subsection{Proof of connectedness of the space of compatible adjacency matrices}
\begin{proof}
Consider two compatible adjacency matrices $\a_{1}$ and $\a_{2}$.
Their Hadamard product $\a_{1}\circ\a_{2}$, which corresponds
to the maximum common subgraph of $\a_{1}$ and $\a_{2}$, is also
compatible. We can remove the edges in $\a_{1}$ that do not appear
in $\a_{1}\circ\a_{2}$ one at a time, and thus arrive at state $\a_{1}\circ\a_{2}$.
Then we add the edges in $\a_{2}$ that do not appear in $\a_{1}\circ\a_{2}$
once at a time, and thus finally arrive at state $\a_{2}$. 
All intermediate adjacency matrices in this procedure
are compatible.
\end{proof}
\subsection{Proof of Theorem \ref{thm:The-proposal-ratio}}
\begin{proof}
In fact, the total number of possible edges that can be added for
$\a(\jmath)$ is given by $\mathrm{Add}(\a(\jmath))$ and the total
number of possible edges that can be removed for $\a(\jmath)$ is
given by $\mathrm{Remove}(\a(\j))$. Thus we have the proposal distribution
\[
\p(\tilde{\a}(\jmath+1)|\a(\jmath))=\frac{1}{\mathrm{Add}(\a(\jmath))+\mathrm{Remove}(\a(\jmath))}.
\]
Similarly, we have
\begin{multline}
\p(\a(\jmath)|\tilde{\a}(\jmath+1))=\\
\frac{1}{\mathrm{Add}(\tilde{\a}(\jmath+1))+\mathrm{Remove}(\tilde{\a}(\jmath+1))}.
\end{multline}
Thus we obtain the proposal ratio:
\begin{multline*}
\frac{\p(\a(\jmath)|\tilde{\a}(\jmath+1))}{\p(\tilde{\a}(\jmath+1)|\a(\jmath))}\\
=\frac{1/\text{\ensuremath{\left(\mathrm{Add}(\tilde{\a}(\jmath+1))+\mathrm{Remove}(\tilde{\a}(\jmath+1))\right)}}}{1/\left(\mathrm{Add}(\a(\jmath))+\mathrm{Remove}(\a(\jmath))\right)}\\
=\frac{\mathrm{Add}(\a(\jmath))+\mathrm{Remove}(\a(\jmath))}{\mathrm{Add}(\tilde{\a}(\jmath+1))+\mathrm{Remove}(\tilde{\a}(\jmath+1))}.
\end{multline*}

\end{proof}

\subsection{Proof of Theorem \ref{thm:likelihood_ratio}}
\begin{proof}
Let $\J^{ab}$ denote the $n\times n$ matrix whose $(a,b)$-entry
is one and all other entries are zeros. We have 
\[
\tilde{\a}(\j+1)-\a(\j)=\pm\left(\J^{ab}+\J^{ba}\right),
\]
where we assign the minus sign ``$-$'' in ``$\pm$'' if the edge that connects subjects
$a$ and $b$ is added to $\tilde{\a}(\j+1)$  and we assign the plus sign ``$+$'' in ``$\pm$''
if that edge is removed from $\a(\j)$. And $e^{\beta(\a)}$ can be
re-written as
\begin{align}
 & e^{\beta(\a)}\nonumber \\
= & \b'\u+\lt(\a\b)'\cdot\mathbf{1}\nonumber \\
= & \mathbf{B}'(\dd-\a\cdot\mathbf{1})+\lt(\a\b)'\cdot\mathbf{1}\nonumber \\
= & \b'\dd-\b'\a\cdot1+\lt(\a\b)'\cdot\mathbf{1}\nonumber \\
= & \b'\dd-\b'\a'\cdot1+\lt(\a\b)'\cdot\mathbf{1}\label{eq:sym}\\
= & \b'\dd-(\a\b)'\cdot1+\lt(\a\b)'\cdot\mathbf{1}\nonumber \\
= & \b'\dd-\ut(\a\b)'\cdot\mathbf{1},\nonumber 
\end{align}
where in (\ref{eq:sym}) we use the fact that $\a$ is symmetric and
$\ut(\cdot)$ denotes the strictly upper triangular part (diagonal
entries exclusive) of a matrix. Thus 
\[
e^{\beta(\a)\pm\J^{ab}}-e^{\beta(\a)}=\pm\ut(\J^{ab}\b)'\cdot\mathbf{1}.
\]
The $(i,j)$-entry of $\J^{ab}\b$ is
\begin{align*}
 & (\J^{ab}\b)_{ij}\\
= & \sum_{k=1}^{n}\J_{ik}^{ab}\b_{kj}\\
= & \J_{ib}^{ab}\b_{bj}.
\end{align*}
Thus
\begin{align*}
 & \left(\ut(\J^{ab}\b)'\cdot\mathbf{1}\right)_{j}\\
= & \sum_{i:i<j}\J_{ib}^{ab}\b_{bj}\\
= & \b_{bj}1\{a<j\}.
\end{align*}
Hence we have
\begin{multline}
(e^{\beta(\tilde{\a}(\j+1))}-e^{\beta(\a(\j))})_{j}\\
=\pm\left(\b_{bj}1\{a<j\}-\b_{aj}1\{b<j\}\right).\label{eq:beta}
\end{multline}
Similarly, if we view $\delta$ in Theorem \ref{thm:likelihood} as
a function of the adjacency matrix $\a$, denoted by $\delta(\a)$,
we have
\begin{multline}
\delta(\tilde{\a}(\j+1))-\delta(\beta(\a(\j)))\\
=\pm\left(\D_{bj}1\{a<j\}-\D_{aj}1\{b<j\}\right).\label{eq:delta}
\end{multline}
If we plug in Equations (\ref{eq:beta}) and (\ref{eq:delta}) in the likelihood
ratio $\frac{\Lambda_{\gamma_{\j}}(\tilde{\a}(\j+1);{\tt,\hat{\theta}_{\iota})}}{\Lambda_{\gamma_{\j}}(\a(\j);\tt,\hat{\theta}_{\iota})}$,
we obtain the expression in the theorem statement.\end{proof}

\end{document}